\documentclass{aa}
\usepackage{graphicx}
\begin{document}
   \title{Implications of the central metal abundance peak in cooling
   core clusters of galaxies}


   \author{H. B\"ohringer\inst{1}, K. Matsushita\inst{2},E. Churazov\inst{3}, A. Finoguenov\inst{1},
          Y. Ikebe\inst{4}
          }

   \offprints{H. B\"ohringer}

   \institute{$^1$ Max-Planck-Institut f\"ur extraterr. Physik,
                 D 85740 Garching, Germany\\
              $^2$ Department of Physics, Tokyo University of Science,
                 1-3 Kagurazaka, Shinjyuku, Tokyo 162-8601, Japan\\
              $^3$ Max-Planck-Institut f\"ur Astrophysik,
                 D 85740 Garching, Germany \\
              $^4$ Joint Center for Astrophysics, Physics Department,
                 University of
                 Maryland, Baltimore County, 1000 Hilltop Circle,
                 Baltimore, MD 21250, USA\\
             }

   \date{Received .....; accepted .....}

   \abstract{Recent XMM-Newton observations of clusters of galaxies
   have provided detailed information on the distribution of 
   heavy elements in the central regions of clusters with cooling
   cores providing strong evidence that most
   of these metals come from recent SN type Ia. In this
   paper we compile information on the cumulative mass profiles of
   iron, the most important metallicity tracer. We find that
   long enrichment times ($\ge 5$ Gyr) are necessary to produce the central
   abundance peaks. Classical cooling flows, a strongly convective 
   intracluster medium, and a complete metal mixing by cluster mergers
   would destroy the observed abundance peaks too rapidly. Thus the
   observations set strong constraints on cluster evolution models
   requiring that the cooling cores in clusters are preserved over
   very long times. We further conclude from the observations that the
   innermost part of the intracluster medium is most probably dominated by 
   gas originating predominantly from stellar mass loss of the cD galaxy.}
  
  \authorrunning{B\"ohringer et al.}
  \titlerunning{Implications of Central Metal Abundance Peaks in Galaxy Clusters}
   \maketitle
%

\section{Introduction}

Since the discovery of the iron emission line in the thermal spectra
of the intracluster Medium (ICM) in clusters of 
galaxies (Mitchell et al. 1976), the
abundances of the heavy elements in the ICM received much attention
and provided insights into the star formation
history in the cluster volumina. The large amount of iron implied that
the cluster galaxies must have lost a large fraction of the iron
produced during their star formation histories to the ICM.
The total iron mass in the ICM is in fact larger than
the total iron mass in all cluster galaxies. Attempts to model the
production of the observed iron masses on the basis of the observed
present day stellar population have shown that such models have to
make maximizing assumptions like a top heavy IMF in early epochs of
star formation to increase the number of historic SN II 
(Arnaud et al. 1992, Elbaz et al.  1993) or an increased rate
of type Ia supernovae in the past (Renzini et al. 1993). 

In this paper we concentrate on the implications of the heavy element
abundances in the central regions. With the ASCA satellite observatory
which provided the first possibility of spatially resolved
spectroscopy of the cluster ICM it was found that some clusters with
cD galaxies had a strong increase in the metal abundance towards the
center (Fukazawa et al. 1996, 2000, Matsumoto et al. 1996, Matsushita 
et al. 1998, Ikebe et al. 1999). With similar results from BeppoSAX
DeGrandi and Molendi (2001)  and DeGrandi et al. (2003)
showed that these iron abundance peaks are a signature of so-called
cooling flow clusters, cluster with a very peaked X-ray surface
brightness profile and thus a high central gas density, which we term
cooling core clusters in this paper.
Again with observational data from ASCA, Finoguenov et al. (2000) could
derive abundance profiles of Fe and Si and could show that the central
regions in those clusters with metal abundance peaks where strongly
enriched in iron while the bulk volume of the clusters outside the
central region had an iron-to-silicon ratio close to the yields of SN
type II. This implies that the central excess seen in cooling core
clusters is most probably due to enrichment by SN type Ia in the cD
galaxies.

Now with the improved capabilities of the XMM-Newton satellite
observatory as well as with the Chandra observatory, more
detailed information on the origin, the metal source distribution, and the
ICM transport history can be obtained. In this paper we compile
information on the central element abundance profiles from four
nearby clusters observed with XMM and interpret the results. 
We discuss the origin of the X-ray emitting gas in terms of the 
enrichment by supernovae type Ia and whether it should be considered
to be the interstellar medium of the 
central galaxy or the cluster ICM. We calculate the time it takes 
to produce the Fe abundance peak and discuss the implications of
the implied large enrichment times for the cluster evolution history.

Throughout the paper we
use a Hubble constant of $h_{70} = 1$ where $h_{70} = H_0 / 70$ km
s$^{-1}$ Mpc$^{-1}$. The other cosmological parameters play not
significant role at the relevant distances. The Virgo distance is
assumed independently of $H_0$ as $17$ Mpc.            

\section{Origin of the metal abundance peak}

   \begin{figure}
   \centering
\includegraphics[width=8.2cm]{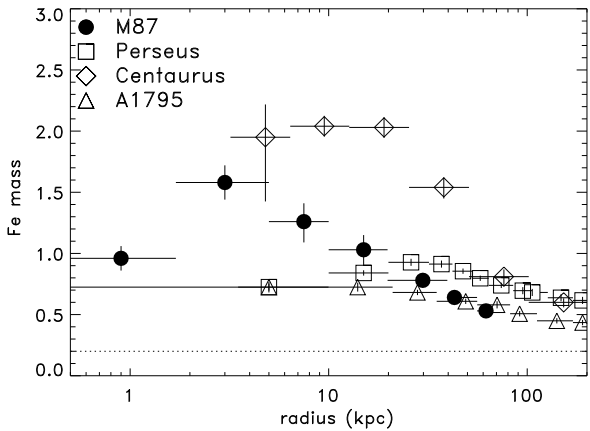}
      \caption{ Iron abundance profiles measured in the four nearby
galaxy clusters, M87/Virgo, Perseus, Centaurus, and A1795. The data are taken
from Matsushita et al. (2003), Churazov et al. (2003), Matsushita et al. (2004),
Tamura et al. (2001), respectively. The values are in solar units based on
the solar abundance of iron quoted by Feldman (1992). The dashed line shows 
the iron abundance with a value of $\sim 0.2$ solar, 
observed on large scale in clusters and assumed to come mostly from SN II 
enrichment before cluster formation. 
}
         \label{Fig1}
   \end{figure}

Recent XMM-Newton observations of M87 (B\"ohringer et
al. 2001, Matsushita et al. 2002, 2003a, Molendi \& Gastaldello 2001,
Gastaldello \& Molendi 2002) have yielded
radial abundance profiles for the M87 halo region in the center of the
Virgo cluster of the seven elements, O, Mg, Si, S, Ar, Ca, Fe. This
provides a finger print of the origin of the metals. Since O and Mg
are essentially only produced by SN II, the observed flat central O abundance
profile is a strong further confirmation of the dominant SN Ia
contribution to the central abundance peak. In an analysis by
Finoguenov et al. (2002) using in one approach e.g. the oxygen
abundance profile to separate the SN II and SN Ia products it was
shown that SN II contributed only about  10\% to the central iron
abundance, corresponding to abundance values of 0.1 to 0.15 in solar units.
The contribution by SN Ia yields an Fe abundance of at least 1 in
solar values.
The SN II abundance contribution in the center of Virgo can also be
compared to the iron abundances in the outer regions of clusters,
where SN II enrichment dominates (Finoguenov et al. 2000, 
DeGrandi \& Molendi 2001) with typical iron abundances of  
about 0.2 of the solar value. Here some contribution 
comes also from SN Ia.
Therefore, we conclude that very roughly a fairly homogeneous
distribution of iron with an abundance of at least 0.15 solar prevails in
these clusters. 

\section{Iron mass contained in the abundance peak}

To further explore the origin of the central abundance peaks we
calculate the total iron mass profiles, $M_{\rm Fe}(\le r)$, for the
clusters Virgo, Centaurus, Perseus, and A1795 from the gas mass
profiles and the Fe abundance profiles as given by Matsushita et
al. (2003, 2004), Churazov et al. (2003), Ikebe et al. (2003;
see also Tamura et al. 2001), respectively. These abundance profiles
for the central 200 kpc are shown in Fig. 1 with abundances given in
solar units based on the solar Fe abundance quoted by Feldman (1992).
Similar results are also reported from the analysis of Chandra
observations showing very good consistency with the results used here
(Sanders et al. 2003, Sanders \& Fabian 2002, and Ettori et al. 2002).
For the Virgo and Centaurus cluster we use the 
abundances derived for the multi-temperature models. The higher 
abundances found for the two- and three temperature modeling only matter
for the very central regions in these two clusters. The gas mass profiles
were also compared to other previous results (e.g. Reiprich \& B\"ohringer
2002) and were found to be in good agreement.

   \begin{figure}
   \centering
\includegraphics[width=8.5cm]{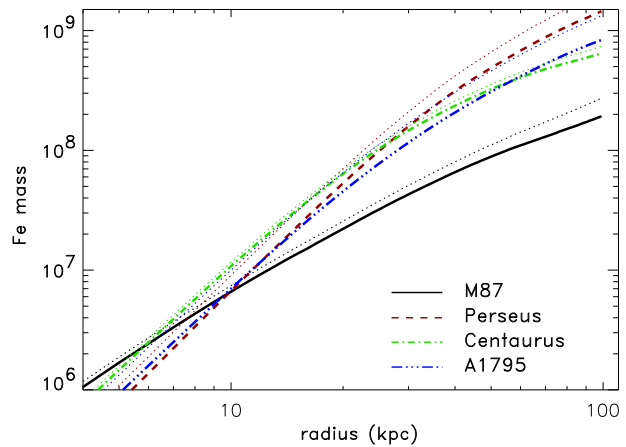}
      \caption{Cumulative iron mass profiles in the centers of four nearby
galaxy clusters. The thin dotted lines show the total iron mass and 
the thick lines show the iron
mass in the central excess where a mean abundance of 0.15 solar was subtracted. 
}
         \label{Fig1}
   \end{figure}

The results in Fig. 2 show as thin dotted
lines the total Fe mass profiles and as solid lines the profiles
for the iron abundance excess, where an ubiquitous Fe abundance of 0.15
of the solar value was subtracted. 
As shown in Fig. 2 the widely distributed Fe component (which we assume to come
from an early enrichment by mainly SN II mostly before the formation of the 
cluster)
is truly a minor fraction inside a radius of 50 kpc. 
(If the very central ICM inside 10 kpc
originates only from stellar mass loss as argued below, the thick lines
underestimate the recently produced Fe mass by a very small amount.)
We note that the iron masses in the abundance excess are quite large. Inside 10 kpc, a radius
that is for example of the order of the scale radius of the giant elliptical galaxies, 
we find that the Fe masses span a narrow range from about 6 to $10\times 10^6$ M$_{\odot}$. 
At a radius of 50 kpc, where in terms of the total 
stellar light the central galaxy is still completely dominant, M87 features a smaller
Fe mass of about $10^8$ M$_{\odot}$ while the more massive clusters show values in the range
$3$ to $5~10^8$ M$_{\odot}$. The smaller Fe mass in the M87 halo is due to 
the smaller total gas mass concentrated around M87 compared to that of
the larger clusters. Also M87 is the least luminous central galaxy in this sample.

   \begin{figure}
   \centering
\includegraphics[width=8.4cm]{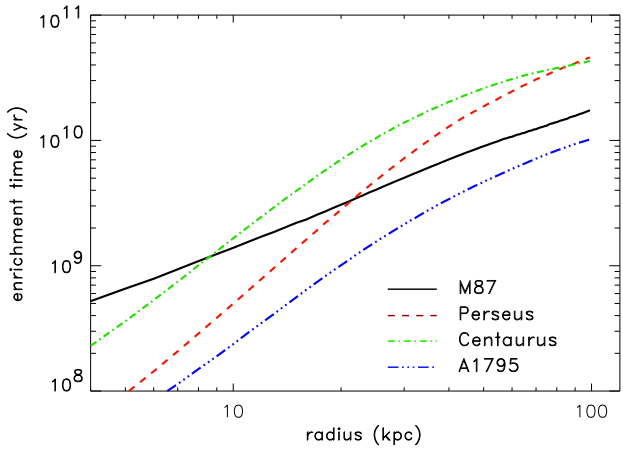}
      \caption{Enrichment ages of the central iron abundance peak 
in the four galaxy clusters, M87/Virgo, Centaurus,
Perseus, and A1795. The enrichment times were calculated 
on the basis of a SN Ia rate of 0.15 SNU
(Capellaro et al. 1999) and an additional contribution by stellar mass loss. 
}
         \label{Fig1}
   \end{figure}

\section{ISM or ICM ?}

It may be to some extent semantic to distinguish the interstellar medium of the central galaxy from
the intracluster medium, since in X-ray images we observe one continuous hot gaseous halo trapped in
the common potential well of the central galaxy and the cluster. It has a very smooth surface 
brightness distribution and temperature profile showing no break, which could lead us
to draw a boundary between two distinct media. Therefore the
distinction has to be made on the basis of the composition. 

We take M87, which has by far the best data, as a study case and concentrate
on the inner 10 kpc. We find a total gas mass of about
$2 \times 10^9$  M$_{\odot}$, which can be replenished by stellar mass loss, taking a rate
as proposed by Ciotti et al. (1991, of about $2.5
\times 10^{-11}$ L$_B$ - assuming a galactic age of 10 Gyr)
in about 2-3 Gyrs. Also the total iron mass of about
$6 \times 10^6$  M$_{\odot}$ can be produced by the SNIa rate of 0.15
SNU (Cappelaro et al. 1997, 1999)
and stellar mass loss from the stellar population of M87 (contributing about 40\%)
in this time. Furthermore as shown by Matsushita et al. (2003) the Mg abundance in the
very center approaches that of the stellar population. Taking these three facts together
the gaseous halo of M87 and its composition in the inner 10 kpc can very well be explained
by stellar mass loss, that is an ISM nature, if it has been produced without disturbance
in the last 2-3 Gyrs. The cooling time of this gas is less than 2 Gyrs, however, and
the mass loss by a traditional cooling flow would be an important sink for this medium
with losses faster than the replenishment by stellar mass loss. Since this traditional
cooling flow picture with large mass deposition rates is not supported by the recent
XMM-Newton and Chandra observations (e.g. Tamura et al. 2001, Peterson et al. 2003,
B\"ohringer et al. 2002), this also implies that the gas is not cooling at
the maximal allowed cooling rate. Therefore the gas in the central
region can be considered as ISM and the neighbouring region outside as
a mixture of ISM and ICM.

   \begin{figure}
   \centering
\includegraphics[width=8.8cm]{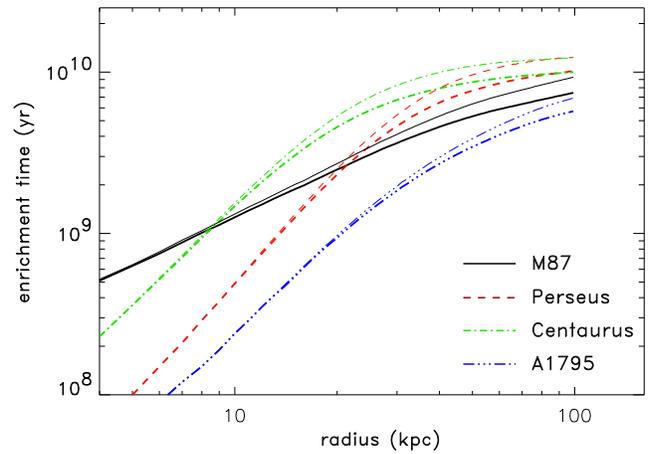}
      \caption{Enrichment ages of the central iron abundance peak 
in the four nearby galaxy clusters for the case of a steeply 
increasing SN type Ia rate in the past (the thick lines are for an 
evolutionary exponent $s=2$ and the thin lines for $s=1.1$.).
}
         \label{Fig1}
   \end{figure}

\section{Enrichment ages for cD halos in galaxy clusters}

For the further understanding of the enrichment and transport processes in these environments
we introduce the concept of enrichment time of the central ICM. 
We take the excess iron abundance as
the enrichment tracer.
We assume that Fe is introduced into the ICM by input from SNIa explosions and
by stellar mass loss from a stellar population heavily 
enriched in Fe. 
The enrichment time, $t_{\rm enr}$, 
is calculated by means of the formula
$$ t_{\rm enr} = M_{\rm Fe} \times \left[ L_{\rm B}~ \left( SR~ 10^{-12} 
{\rm L}_{\rm B\odot}^{-1} \eta_{\rm Fe} + 2.5 \times 10^{-11} 
{\rm L}_{\rm B\odot}^{-1} \gamma_{\rm Fe} \right) \right] ^{-1} $$
where $L_{\rm B}$ is the galaxy blue luminosity inside $r$,
$SR$ is the SN Ia rate in SNU, $\eta_{\rm Fe}$ is the iron
yield per SN Ia (0.7 M$_{\odot})$, and $\gamma_{\rm Fe}$
is the mean iron mass fraction in the stellar winds
($2.8 \times 10^{-3}$) for a mean stellar iron abundance of about 1.6.
The light profile of the central galaxies was obtained from
Giraud (1999) for M87, from Schombert (1988) for NGC1275, and we used
the generalized model from Schombert (1986) together with the 
total luminosity for the central galaxies of Centaurus (NGC 4696)
and A1795. The value for the total luminosity of A1795 was taken from 
the I-band observation of Johnstone et al. (1991).

Fig. 3 shows the derived enrichment
times. While inside 10 kpc the enrichment times are typically
less than 2 Gyrs, the times become quite large at the
50 kpc radius. While for A1795 and M87 we find about 5 and 9
Gyrs the enrichment times exceed the Hubble time for 
Centaurus and Perseus. The relatively short time for A1795 
comes as a consequence in the calculations 
of the very luminous central galaxy. 

Alternatively, we can adopt an increased SN Ia rate, $R(t)$, in
the past as assumed in e.g. Renzini et al. (1993), often modeled
by a power law behavior
$$R(t) = R_0 ~\left(t\over t_0\right)^{-s}$$
We explore two models with $s= 1.1$ and $2$ ($t_0 = 13$ Gyr).
The latter model has a more extreme exponent than used
in  Renzini et al. which is here
meant to represent a bracketing solution.
The so determined enrichment
times are shown in Fig. 4.  The most important contribution 
now is happening at the earliest times, which
results in a flattening of the curves. At a radius of 50 kpc the
different examples converge to a narrow age span {of 5 -
9 (6.4 - 11) Gyr, with the exception of A1795 which yields 
about 3.5 (3.9) Gyrs for the $s=2$ ($s=1.1$) model.
We therefore conclude that unless the supernova rate for SN Ia increases
in the past we cannot explain the 
production of the iron peak, but we do net get further constraints
for the parameter $s$ other than it needs to be at least close to one. 
} 

{DeGrandi et al. (2003) also discuss in detail the origin of iron 
abundance peaks in the centers of clusters, taking a 
different look at this phenomenon, however. While we have concentrated
here on the inner region of the abundance peak and on
the recent enrichment by SN Ia (motivated by the abundance pattern
described in section 2), these authors describe
the total iron mass of the central abundance excess
streching over a range of at least 200 - 300 kpc. Consequently they
find higher values for the excess iron mass, typically
$ 1 - 4 \times 10^9$ M$_{\odot}$. To explain the creation of 
this abundance peak, they investigate
the total iron ejection of the central galaxy over its lifetime
based on the modeling of Pipino et al. (2002) including both supernovae
type II and type Ia and find that the total iron production is sufficient
to account for the observed central excess. This is covering a 
much larger time span including the early epoch of SN II enrichment
which lays mostly before the formation of the cluster.
It also implies a different element composition for the large scale
abundance peak than we find for the innermost part.}

\section{Implications for cluster cooling flows} 

Turning again to the classical cooling flow picture, we can with the
given observed parameters and with the known mass deposition profiles
calculated in the classical way for cooling flow models (e.g. Allen et
al. 2001) determine the iron mass loss rate from the ICM due to mass
deposition for a steady state situation and the given abundances. { The
derived iron depletion times inside a radius of 50 (10) kpc are 
smaller by factors of 1 (2), 6 (1), 6.5 (8), 1.5 (0.3) compared to
the enrichment times for 
M87, Perseus, Centaurus, and A1795, respectively.}
Therefore it seems very difficult to sustain the very
large central iron concentrations in the frame of the classical 
cooling flow model with large mass deposition rates.
Reduced mass deposition rates suggested by recent observations
(e.g. Petersen et al. 2003) would not strongly affect the above
enrichment picture, however.

\section{Implications for other ICM transport histories}

The long enrichment times also
put constraints on any disturbances of the
central regions of the clusters. A first disturbance could be provided
by the proposed heating models of cooling core regions by AGN. If
e.g. the heating model by bubbles of relativistic plasma from radio
lobes, as proposed by Churazov et al. (2000, 2001), leads to a large
convective motion covering the full cooling core region, gas could be
transported around in a circle in a few Gyr. This would tend to
flatten the observed steep abundance profiles. Numerical simulations
as performed e.g. by Br\"uggen (2002) where the effect of buoyantly
rising bubbles was studied in the presence of tracer particles show
that the convection induced in the ambient gas by the relativistic
plasma bubbles may be more ``differential'', such that individual gas
parcels are only dragged along for shorter distances. Only such a gentle
motion would be consistent with the observed gas profiles. 
Other proposed heating mechanisms for cooling cores, 
not relying on convection, like heating by viscous sound or weak shock
waves (Fabian et al. 2003, Ruszkowski et al. 2003, Forman et al. 2003)
would not affect the steep observed abundance gradients.

Finally the observations also put strong constraints on the effects of
cluster mergers on the cooling cores of clusters. Traditionally
cluster mergers are believed to destroy cooling flows 
(e.g. McGlynn \& Fabian, 1984). Allen et al. (2001) has calculated 
cooling flow ages on the basis of the comparison of the spectroscopically
inferred mass flow rates with those rates determined from the surface 
brightness distribution and finds cooling flow ages of the order of
a few to several Gyr. This is consistent with the merger frequency 
expected to be a few Gyr (e.g. Schuecker et al. 2001). The above analysis
suggests that the cooling cores are in general older than the time to the
last major merger. This implies that most mergers may not destroy 
the cooling cores. Specificly, more rare merger configurations with 
roughly equal mass partners and both with dense cores may be required 
for an effective cooling core disruption and the cooling core clusters
we see today may never have suffered from such an event, while clusters
like e.g. Coma have. The difficulty of disrupting the cluster core by
a merger was also pointed out by Churazov et al. (2003) in
a different context. For a formation scenario for cooling cores 
which is alternative to cooling flow models see 
also Motl et al. (2003). The enrichment ages also provide
a lower limit on the age of the central region of
the cluster, but a definitive age estimate has to await a more precise
picture of the enrichment history to develop.

\section{Summary and conclusions}

The large metal abundance excesses in cooling core regions of 
clusters of galaxies indicate that large enrichment times are necessary
to create them. For the four examples studied here, the Virgo, Perseus,
Centaurus, and Abell 1795 clusters, we find enrichment times of about 4 - 10
Gyrs for the central 50 kpc radius region. Thus we conclude that the 
dense cooling cores in clusters are very persistent phenomena. 
These studies can easily be expanded to a larger sample of clusters to test
if these conclusions hold in general. But since very similar abundance
gradients are observed for most cooling core clusters we expect that this 
will be the case. A more detailed study of the abundance ratios of several
tracer elements will also provide more detailed information on the 
enrichment history by the different SN types as well as on the possible 
secular evolution of the SNIa.

\begin{acknowledgements}
The paper is based on data obtained with XMM-Newton, an ESA
science mission with instruments and contributions directly funded by
ESA Member States and the USA (NASA). The XMM-Newton project is supported
by the Bundesministerium f\"ur Bildung und Forschung, Deutsches Zentrum
f\"ur Luft und Raumfahrt (BMBF/DLR), the Max-Planck Society and the
Haidenhain-Stiftung.

\end{acknowledgements}

\end{document}